\documentclass{revtex4}

\usepackage{graphicx}
\usepackage{bm}
\usepackage{mathrsfs}

\newcommand{\be}{\begin{equation}}
\newcommand{\ee}{\end{equation}}
\newcommand{\ba}{\begin{eqnarray}}
\newcommand{\ea}{\end{eqnarray}}
\newcommand{\bra}[1]{\left(#1\right)}
\newcommand{\bras}[1]{\left[#1\right]}
\newcommand{\brac}[1]{\left\{#1\right\}}

\renewcommand{\:}[2]{{\textstyle\frac{#1}{#2}}}
\renewcommand{\;}[2]{{\frac{#1}{#2}}}

\newcommand{\forget}[1]{\iffalse#1\fi}
\newcommand{\forgetmenot}[1]{\iftrue#1\fi}

\newcommand{\del}{\nabla}

\newcommand{\curl}{{\mathsf{curl}\,}}
\newcommand{\sdel}{{\mathrm{D}}}

\newcommand{\<}{\langle}
\renewcommand{\>}{\rangle}

\newcommand{\uudot}{\dot{u}}
\newcommand{\udot}{{\cal A}}
\newcommand{\n}{n}
\newcommand{\N}{N}
\newcommand{\E}{{\cal E}}
\renewcommand{\H}{{\cal H}}
\newcommand{\lc}{\varepsilon}
\newcommand{\hatn}{a}
\newcommand{\dotn}{\alpha}
\newcommand{\lb}{\{}
\newcommand{\rb}{\}}

\newcommand{\tfrac}{\:}
\newcommand{\El}{\mathscr{E}}
\newcommand{\B}{\mathscr{B}}

\newcommand{\T}{_{\mathsf{T}}}

\newcommand{\Si}{{\mathsf{S}}}
\newcommand{\Vi}{{\mathsf{V}}}

\begin{document}

\title{A covariant approach for perturbations of rotationally symmetric spacetimes}
\author{Chris Clarkson}
\affiliation{Cosmology and Gravity Group, Department of Mathematics and Applied Mathematics, University of Cape Town, Rondebosch 7701, Cape Town, South Africa}
\email{chris.clarkson@uct.ac.za}

\date{\today}

\begin{abstract}

We present a covariant decomposition of Einstein's Field Equations which is particularly suitable for perturbations of
spherically symmetric -- and general locally rotationally symmetric
-- spacetimes. Based upon the utility of the 1+3 covariant
approach to perturbation theory in cosmology, the semi-tetrad,
1+1+2 approach presented here should be useful for analysing perturbations
of a variety of systems in a covariant and gauge-invariant manner. Such applications range from stellar objects to cosmological models such as the spherically symmetric Lema\^\i tre-Tolman-Bondi solutions or the class of locally rotationally symmetric Bianchi models. 

\end{abstract}
\maketitle

\section{introduction}

Tetrad formalisms in general relativity have played a pivotal role
in its development as well as our understanding of the subject.
These range from the complex null tetrad of Newmann and Penrose, to the 1+3 approach of Ehlers, Ellis and others, which includes both a full tetrad approach as well as a partial `covariant' approach where only one timelike tetrad vector is chosen  (see~\cite{HvE} for a review and references). These techniques
formulate the equations of general relativity as first-order
differential equations in the physical curvature and dynamic
variables of the covariant derivatives of the tetrad vectors, as opposed to the more usual coordinate approach involving second-order partial differential equations in
functions appearing in the metric. The differential operators which appear are convective derivatives along the tetrad vectors as opposed partial derivatives with respect to particular coordinates. Much of their utility arises in
spacetimes with special symmetry. For example, the 1+3 covariant approach is
perfect for cosmology because it covariantly  factorises out the
essential coordinate -- time -- leaving all the background field equations as
covariant scalar equations. Under perturbations all 3-vectors and
tensors (which must vanish in the background due to homogeneity and isotropy)
become gauge-invariant first-order quantities making a Fourier analysis easy~\cite{BE,HvE}.

We formulate here an approach which involves a semi-tetrad: we keep the timelike threading vector field of the 1+3 approach and introduce one spatial vector. The remaining two dimensions are left untouched, rather like the `3' in the 1+3 approach. Indeed the formalism presented here may be considered as half-way between the 1+3 tetrad and covariant approaches. A similar approach has been discussed before in~\cite{greenberg,TM1,TM2,T}, and we expand on this considerably here by presenting the full system of 1+1+2 equations.

It is expected that this approach may find use in perturbations of spacetimes with a preferred spatial direction at each point -- so-called locally rotationally symmetric spacetimes~\cite{vEE}. These include the spherically symmetric Lema\^\i tre-Tolman-Bondi models, many classes of Bianchi models, as well as forming the background for most stellar models. In this paper we provide the algorithm of how to calculate gravitational perturbations in any LRS spacetime in a covariant and gauge-invariant (GI) way. 
 
 As an example of it's utility, such a covariant perturbative scheme was applied to the Schwarzschild solution in~\cite{CB}. Despite being a well understood problem, it was shown using the 1+1+2 approach how both the axial and polar degrees of freedom may be unified into a single transverse traceless tensor which obeys the tensorial form of the
Regge-Wheeler equation \cite{RW,CB}
\ba
-\ddot W_{ab}+\hat{\hat W}_{ab}+\udot{\hat W}_{ab} -\phi^2 W_{ab}+\delta^2
W_{ab} =0,\label{RW}
\ea
where the Regge-Wheeler tensor $W_{ab}$ is a gauge- and
frame-invariant TT tensor, defined in~\cite{CB} (other variables are defined below), and $\dot{~},\hat{~}$ and $\delta$ are time, radial and angular derivatives, respectively.
This tensor contains in compact
form the curved space generalisation of the two flat space GW polarizations
$h_+$ and $h_\times$~\cite{CMBD} (see also~\cite{BL2} for an extension of this work). 
The approach here also been used to study scalar and electromagnetic perturbations of LRS spacetimes, and generalised Regge-Wheeler equations were found~\cite{BC,BL1}. Furthermore, it has been used to study the interaction of magnetic fields and gravitational waves around a black hole -- a process which produces electromagnetic radiation mirroring the gravitational waves~\cite{CMBD}.

In Sec.~\ref{formalism} we discuss the 1+1+2 approach in full generality, and then in Sec. III we discuss the perturbation procedure for LRS spacetimes, before sumarising in Sec. IV.

\section{Formalism}\label{formalism}

In the 1+3 approach, a timelike threading vector field $u^a$ ($u^au_a=-1$) is
introduced, representing the observers' congruence. Given this vector field,
the projection tensor $h_{a}^{~b}=g_{a}^{~b}+u_au^b$ is introduced, which
projects all vectors and tensors orthogonal to~$u^a$. Using $h_{ab}$, any
4-vector may be split into a (1+3 scalar) part parallel to $u^a$ and a
(3-vector) part orthogonal to $u^a$. Any second rank tensor may be covariantly
and irreducibly split into scalar, vector and{projected, symmetric,
trace-free (PSTF)} 3-tensor parts, which requires the alternating tensor
$\lc_{abc}=u^d\eta_{dabc}$~\cite{HvE}. Tensors of higher rank may be similarly
split, but are rarely used (an important exception being cosmic microwave background
physics~\cite{CL,MGE}). These are the fundamental quantities describing the
spacetime, after the introduction of~$u^a$.

We now introduce another vector field and perform {another} split, but
this time of the 1+3 equations. The `1+1+2' decomposition we develop here has
been partially studied before, mostly in the context of symmetries of solutions
of the EFE~\cite{zafiris,TM2}. It was introduced by~\cite{greenberg} and
further developed in~\cite{TM1,BC,CB,BL1,BL2}. However, there are importances differences
with the work presented here. {In the following we assume the 1+3
covariant split of the equations} (as given in~\cite{HvE}, for example),
{with all tensors split into scalars, vectors and PSTF tensors with
respect to~$u^a$.}

Take a unit vector~$\n^a$ orthogonal to $u^a$: $\n^a\n_a=1,~u^a\n_a=0$, and
define the projection tensor
\be
\N_a^{~b}\equiv h_a^{~b}-\n_a\n^b=g_{a}^{~b}+u_au^b-\n_a\n^b,
\ee
which projects vectors orthogonal to $\n^a$ (and $u^a$):
$\n^a\N_{ab}=0=u^a\N_{ab}$, onto 2-surfaces ($\N_a^{~a}=2$) which we refer to
as the sheet. This is also the screen space of the null vector
$k^a\propto u^a+\n^a$.

Any 3-vector $\psi^a$ can now be irreducibly split into a scalar, $\Psi$, which
is the part of the vector parallel to $\n^a$, and a vector, $\Psi^a$, lying in
the sheet orthogonal to $\n^a$;
\be
\psi^a=\Psi\n^a+\Psi^{a},~~~\mbox{where}~~~\Psi\equiv \psi_a\n^a,~~~\mbox{and}~~~\Psi^{a}\equiv
\N^{ab}\psi_b\equiv \psi^{\bar a},
\label{vector-decomp}
\ee
where we use a bar over an index to denote projection with
$\N_{ab}$ on that index. Similarly, any PSTF tensor, $\psi_{ab}$,
can now be split into scalar, vector and tensor (which are PSTF
with respect to $\n^a$) parts:
\be
\psi_{ab}=\psi_{\<ab\>}=\Psi\bra{\n_a\n_b-\:12\N_{ab}}+2\Psi_{(a}\n_{b)}+\Psi_{{ab}},
\label{tensor-decomp}
\ee
where
\ba
\Psi&\equiv &\n^a\n^b\psi_{ab}=-\N^{ab}\psi_{ab},\nonumber\\
\Psi_a&\equiv &\N_a^{~b}\n^c\psi_{bc}=\Psi_{\bar a},\nonumber\\
\Psi_{ ab}&\equiv &\psi_{\lb ab\rb}\equiv
\bra{\N_{(a}^{~~c}\N_{b)}^{~~d}-\:12\N_{ab}\N^{cd}}\psi_{cd}\label{PSTF-TT}.
\ea
We use curly brackets to denote the PSTF with respect to $\n^a$ part of a
tensor. {Note that for 2nd-rank tensors in the 1+1+2 formalism `PSTF' is
precisely equivalent to `transverse-traceless'}. Note also that $h_{\lb
ab\rb}=0$,~$\N_{\<ab\>}=-\n_{\<a}\n_{b\>}=\N_{ab}-\:23h_{ab}$.

We also define the alternating Levi-Civita 2-tensor
\be
\lc_{ab}\equiv\lc_{abc}\n^c = u^d\eta_{dabc}n^c,
\ee
so that $\lc_{ab}\n^b=0=\lc_{(ab)}$, and
\ba
\lc_{abc}&=& \n_a\lc_{bc}-\n_b\lc_{ac}+\n_c\lc_{ab},\\
\lc_{ab}\lc^{cd}&=&\N_a^{~c}\N_b^{~d}-\N_a^{~d}\N_b^{~c},\\
\lc_a^{~c}\lc_{bc}&=&\N_{ab},~~~\lc^{ab}\lc_{ab}=2.
\ea
Note that for a 2-vector $\Psi^a$, $\lc_{ab}$ may be used to form a vector
orthogonal to $\Psi^a$ but of the same length.

With these definitions we may split any object into {scalars,
2-vectors in the sheet, and transverse-traceless 2-tensors, also defined in the
sheet.} These three types of objects are the only objects which appear, after a
complete decomposition. Hereafter, we will assume such a split has been made, and
`vector' will generally refer to a vector projected orthogonal to $u^a$ and
$\n^a$, and `tensor' will generally mean transverse-traceless tensor, defined
by Eq.~(\ref{PSTF-TT}).

There are two new derivatives of interest now, which $\n^a$ defines, for any
object $\psi_{\cdots}^{~~\cdots}$:
\ba
\hat \psi_{a\cdots b}^{~~~~~c\cdots d}&\equiv &
\n^e \sdel_e\psi_{a\cdots b}^{~~~~~c\cdots d},\label{hatdef}\\
\delta_e \psi_{a\cdots b}^{~~~~~c\cdots d}&\equiv & \N_e^{~j}\N_a^{~f}\cdots
\N_b^{~g}\N_h^{~c}\cdots\N_i^{~d}\sdel_j \psi_{f\cdots g}^{~~~~~h\cdots i}.\label{deltadef}
\ea
The hat-derivative is the derivative along the vector field $\n^a$ in the
surfaces orthogonal to $u^a$. This definition represents a conceptual
divergence from 1+3 tetrad approach, in which the basis vectors appear on an
equal footing [i.e.,~with $\del_a$ rather than $\sdel_a$ in
Eq.~(\ref{hatdef})]. As a result, the congruence~$u^a$ retains the primary
importance it has in the 1+3 covariant approach.
(We choose to think of ${\cal A}\equiv u^an^b\del_a u_b= -u^au^b\del_a n_b$ as
the radial component of the acceleration of~$u^a$, rather than the time
component of $\dot{n}^a$.) The $\delta$-derivative, defined by
Eq.~(\ref{deltadef}) is a projected derivative on the sheet, with projection on
every free index.

These derivatives then affect our projection tensor $N_{ab}$ and Levi-Civita tensor as follows:
\ba
\dot\N_{ab}&=&2u_{(a}\uudot_{b)}-2\n_{(a}\dot\n_{b)}=2u_{(a}\udot_{b)}-2\n_{(a}\alpha_{b)},\\
\hat\N_{ab}&=&-2\n_{(a}\hat\n_{b)},\\
\delta_c\N_{ab}&=&0,\\
\dot\lc_{ab}&=&-2u_{[a}\lc_{b]c}\udot^c+2\n_{[a}\lc_{b]c}\dotn^c,\\
\hat\lc_{ab}&=&2\n_{[a}\lc_{b]c}\hatn^c,\\
\delta_c\lc_{ab}&=&0.
\ea

We now decompose the covariant derivative of $\n^a$
orthogonal to $u^a$ into its irreducible form:
\be
\sdel_a\n_b=\n_a\hatn_b+\:12\phi \N_{ab}+\xi\lc_{ab}+\zeta_{ab},
\ee
where
\ba
\hatn_a &\equiv &\n^c\sdel_c\n_a=\hat \n_a,\hspace{10cm}\\
\phi &\equiv &\delta_a \n^a,\\
\xi &\equiv &\:12\lc^{ab}\delta_a\n_b,\\
\zeta_{ab} &\equiv &\delta_{\lb a}\n_{b\rb}.
\ea
We may interpret these as follows: travelling along $\n^a$, $\phi$ represents
the sheet expansion, $\zeta_{ab}$ is the shear of $\n^a$ (distortion of the
sheet), and $\hatn^a$ its acceleration, while $\xi$ represents a `twisting' of
the sheet~-- the rotation of $\n^a$~\cite{TM1}. The other derivative of $\n^a$
is its change along $u^a$,
\be
\dot n_{a}=\udot u_a+\dotn_a~~~\mbox{where}~~~\dotn_a\equiv\dot n_{\bar a}
~~~\mbox{and}~~~\udot=\n^a\uudot_a.
\ee
The new variables $\hatn_a$, $\phi$, $\xi$, $\zeta_{ab}$ and $\dotn_a$ are
fundamental objects in the spacetime, and their dynamics gives us information
about the spacetime geometry. They are treated on the same footing as the
kinematical variables of $u^a$ in the 1+3 approach (which also appear here).

For any vector $\Psi^a$ orthogonal to $\n^a$ and $u^a$
(i.e.,~$\Psi^a=\Psi^{\bar a}$), we may decompose the different parts of its
spatial derivative:
\be
\sdel_a \Psi_b=-\n_a\n_b\Psi_c\hatn^c+\n_a\hat\Psi_{\bar b}-\n_b\bras{\:12\phi\Psi_a+
\bra{\xi\lc_{ac}+\zeta_{ac}}\Psi^c}+\delta_a\Psi_b.
\ee
Similarly, for a tensor $\Psi_{ab}$: $\Psi_{ab}=\Psi_{\{ab\}}$, we have
\be
\sdel_a\Psi_{bc}=-2\n_a \n_{(b}\Psi_{c)d}\hatn^d
 +\n_a\hat \Psi_{bc}-2\n_{(b}\bras{
\:12\phi \Psi_{c)a}+\Psi_{c)}^{~~d}\bra{\xi \lc_{ad}+\zeta_{ad}}}
 +\delta_a\Psi_{bc}.
\ee
Note that for a scalar, we have $\sdel_a\Psi=\hat\Psi\n_a+\delta_a\Psi$.

We take $\n^a$ to be arbitrary at this point, and then split the usual 1+3
kinematical and Weyl quantities into the irreducible
set~$\{\theta,\udot,\Omega,\Sigma,{\cal E},{\cal H},\udot^a,\Sigma^a,{\cal
E}^a,{\cal H}^a,\Sigma_{ab},{\cal E}_{ab},{\cal H}_{ab}\}$ using
(\ref{vector-decomp}) and~(\ref{tensor-decomp}) as follows:
\ba
\uudot^a&=&\udot \n^a+\udot^a,\\
\omega^a&=&\Omega \n^a+\Omega^a,\\
\sigma_{ab}&=&\Sigma\bra{\n_a\n_b-\:12\N_{ab}}+2\Sigma_{(a}\n_{b)}+\Sigma_{ab},\\
E_{ab}&=&{\cal E}\bra{\n_a\n_b-\:12\N_{ab}}+2{\cal E}_{(a}\n_{b)}+{\cal E}_{ab},\\
H_{ab}&=&{\cal H}\bra{\n_a\n_b-\:12\N_{ab}}+2{\cal H}_{(a}\n_{b)}+{\cal
H}_{ab}.
\ea
The shear scalar, $\sigma$, for example, may be expressed in the
form
\be
\sigma^2 \equiv \:12\sigma_{ab}\sigma^{ab} = \:34\Sigma^2 +
\Sigma_a\Sigma^a + \:12\Sigma_{ab}\Sigma^{ab}.
\ee
Similarly we may split the fluid variables $q^a$ and $\pi_{ab}$,
\ba
q^a&=&Q \n^a+Q^a,\\
\pi_{ab}&=&\Pi\bra{\n_a\n_b-\:12\N_{ab}}+2\Pi_{(a}\n_{b)}+\Pi_{ab}.
\ea

Having described the splitting of the 1+3 variables to obtain
their 1+1+2 parts, and the introduction of the new 1+1+2 variables
corresponding to the irreducible parts of~$\del_a\n_b$, it only
remains to apply this decomposition procedure to the 1+3 equations
themselves, as well as the Ricci identities for $n^a$. We give these equations in
section~\ref{the-equations}.

\subsection{Commutation relations}

In general the three derivatives we now have defined,
$`\dot{\phantom{x}}$',~$`\hat{\phantom{x}}$' and $`\delta_a$' do not commute.
Instead, when acting on a scalar $\psi$, they satisfy:
\ba
\hat{\dot \psi}-\dot{\hat \psi}&=&-\udot\dot\psi+\bra{\:13\theta+\Sigma}\hat\psi
+\bra{\Sigma_a+\lc_{ab}\Omega^b-\dotn_a}\delta^a\psi,\label{comm-un}
\\
\delta_a\dot\psi-\N_a^{~b}\bra{\delta_b\psi}^\cdot&=&-\udot_{a}\dot\psi+
\bra{\dotn_{a}+\Sigma_a-\lc_{ab}\Omega^b}\hat\psi
+\bra{\:13\theta-\:12\Sigma}\delta_a\psi +
\bra{\Sigma_{ab}+\Omega\lc_{ab}}\delta^b\psi,\label{commdd}
\\
\delta_a\hat\psi-\N_a^{~b}\widehat{\bra{\delta_b\psi}}
&=&\bra{\Sigma_a-\lc_{ab}\Omega^b}\dot\psi+\hatn_a\hat\psi+\:12\phi\delta_a\psi+
\bra{\zeta_{ab}+\xi\lc_{ab}}\delta^b\psi,\label{commdh}
\\
\delta_a\delta_b\psi-\delta_b\delta_a\psi&=& 2\lc_{ab}\bra{\Omega\dot\psi-\xi\hat\psi}
+2\hatn_{[a}\delta_{b]}\psi.\label{comm1}
\ea
The commutation relations for 2-vectors $\psi_a$ are
\ba
\hat{\dot \psi}_{\bar a}-\dot{\hat \psi}_{\bar
  a}&=&-\udot\dot\psi_{\bar a}+\bra{\:13\theta+\Sigma}\hat\psi_{\bar a}
+\bra{\Sigma_b+\lc_{bc}\Omega^c-\dotn_b}\delta^b\psi_a + \udot_a
\bra{\Sigma_b+\lc_{bc}\Omega^c}\psi^b + \H\lc_{ab}\psi^b
  ,\label{commv-un}
\\
\delta_a\dot\psi_b-\N_a^{~c}\N_b^{~d}\bra{\delta_c\psi_d}^\cdot&=&-\udot_{a}\dot\psi_b+
\bra{\dotn_{a}+\Sigma_a-\lc_{ac}\Omega^c}\hat\psi_{\bar b}
+\bra{\:13\theta-\:12\Sigma}\bra{\delta_a\psi_b+\psi_a\udot_b} +
\H_a\lc_{bc}\psi^c \nonumber \\ && +
\bra{\Sigma_{ac}+\Omega\lc_{ac}}\bra{\delta^c\psi_b+\psi^c\udot_b}
+\:12\bra{\psi_aQ_b-N_{ab}\psi^cQ_c}\nonumber \\ &&
 -\bra{\:12\phi
  N_{ac}+\xi\lc_{ac}+\zeta_{ac}}\psi^c\alpha_b, \label{commv-deldot}
\\
\delta_a\hat\psi_b-\N_a^{~c}\N_b^{d}\widehat{\bra{\delta_c\psi_d}}
&=&\bra{\Sigma_a-\lc_{ac}\Omega^c}\dot\psi_{\bar b} +
\hatn_a\hat\psi_{\bar b}
 +\:12\phi\bra{\delta_a\psi_b-\psi_aa_b}+\bra{\zeta_{ac}+\xi\lc_{ac}}\bra{\delta^c\psi_b-\psi^ca_b}\nonumber \\
&& + N_{ab}\psi^c\bra{\:12\Pi_c+\E_c} - \psi_a\bra{\:12\Pi_b+\E_b},\label{commv-delhat}
\\
\delta_a\delta_b\psi_c-\delta_b\delta_a\psi_c&=&
2\lc_{ab}\bra{\Omega\dot\psi_{\bar c}-\xi\hat\psi_{\bar c}}
+2\bras{\bra{\:13\theta-\:12\Sigma}^2 -\:14\phi^2+\:12\Pi+\E
  -\:13\bra{\mu+\Lambda}}\psi_{[a}N_{b]c} \nonumber \\
&&
-2\psi_{[a}\bras{-\bra{\:13\theta-\:12\Sigma}\bra{\Sigma_{b]c}+\Omega\lc_{b]c}}
+\:12\phi\bra{\zeta_{b]c}+\xi\lc_{b]c}} +\:12\Pi_{b]c}+\E_{b]c}}
      \nonumber \\
&& +2\N_{[ac}\bras{-\bra{\:13\theta-\:12\Sigma}
 \bra{\Sigma_{b]d}+\Omega\lc_{b]d}}
+\:12\phi\bra{\zeta_{b]d}+\xi\lc_{b]d}} +\:12\Pi_{b]d}+ȧ\E_{b]d}}\psi^d
      \nonumber \\
&&+2\bras{-\bra{\Sigma_{[ac}+\Omega\lc_{[ac}}\bra{\Sigma_{b]d}+\Omega\lc_{b]d}}
+\bra{\zeta_{[ac}+\xi\lc_{[ac}}\bra{\zeta_{b]d}+\xi\lc_{b]d}}}\psi^d
          .\label{commv1}
\ea
These relations are more complicated for tensors.
These last two equations in the case of scalars are the decomposition of the 1+3 commutation relation
\be
\curl\sdel_a\psi=2\dot\psi\omega_a.
\ee

From Eq.~(\ref{comm1}), we see that our sheet will be a genuine 2-surface in
the spacetime (and, in particular, that the derivative~$\delta_a$ will be a
true covariant derivative on this surface) if and only if
$\xi=\Omega=\hatn^a=0$. (Recall that the 1+3 spatial metric $h_{ab}$
corresponds to a genuine 3-surface when $\omega^a=0$.) Otherwise, the sheet is
really just a collection of tangent planes. In addition, the two vectors $u^a$
and $\n^a$ are 2-surface forming if and only if the commutator $[u,n]$
in~(\ref{comm-un}) has no component in the sheet: that is, when Greenberg's
vector
\be
\Sigma^a+\lc^{ab}\Omega_b-\dotn^a
\ee
vanishes~\cite{zafiris}~-- see Eq.~(\ref{comm-un}).

\section{The equations} \label{the-equations}

Once the vector $\n^a$ has been introduced it is possible, and necessary, to
augment the 1+3 equations with the Ricci identities for~$\n^a$; without these
we do not have enough equations to determine the new 1+1+2 variables. The Ricci
identities for $\n^a$ are
\be
R_{abc}\equiv2\del_{[a}\del_{b]}\n_c-R_{abcd}\n^d=0,\label{ricci}
\ee
where $R_{abcd}$ is the Riemann curvature tensor. This third-rank tensor may be
covariantly split using the two vector fields $u^a$ and $n^a$, and gives
dynamical equations for the covariant parts of the derivative of $\n^a$ (namely
$\dotn_a$, $\hatn_a$, $\phi$, $\xi$ and~$\zeta_{ab}$) in the form of
\emph{evolution} equations, involving dot-derivatives of these variables, and
\emph{propagation} equations, involving hat-derivatives. In order to facilitate
the calculation of these Ricci identities, which appear in the following
section, we give here the expression for the full covariant derivative of
$\n^a$ in terms of the relevant 1+1+2 variables:
\be
\del_a\n_b=-\udot u_au_b-u_a\dotn_b +\bra{\Sigma+\:13\theta}\n_a u_b
+ \bra{\Sigma_a-\lc_{ac}\Omega^c}u_b +\n_a\hatn_b
+\:12\phi\N_{ab}+\xi\lc_{ab}+\zeta_{ab},
\ee
which may be inserted into Eq.~(\ref{ricci}). The full decomposition of the
covariant derivative of $u^a$ is
\ba
\del_au_b&=&
-u_a\bra{\udot\n_b+\udot_b}+\n_a\n_b\bra{\:13\theta+\Sigma}
+\n_a\bra{\Sigma_b+\lc_{bc}\Omega^c}\nonumber\\&&
+\bra{\Sigma_a-\lc_{ac}\Omega^c}\n_b+\N_{ab}\bra{\:13\theta-\:12\Sigma}+
\Omega\lc_{ab}+\Sigma_{ab},
\ea
which in turn implies the useful relation
\be
\hat{u}_a =
\bra{\:13\theta+\Sigma}\n_a+\Sigma_a+\lc_{ab}\Omega^b.
\label{uhat}
\ee

We have now assembled all the tools necessary to provide the full
system of equations for the 1+1+2 formalism. This consists of
evolution equations, propagation equations, mixtures of both, and
constraints. Formulae which are useful for splitting 1+3 equations are given in Appendix~\ref{split}.

\subsection{Evolution equations}

We find evolution equations for the 1+1+2 variables $\phi, \xi$ and $\zeta_{ab}$  from the projection $u^aR_{abc}$. 

$u^a\N^{bc}R_{abc}$:
\be
\dot\phi =
\bra{\:23\theta-\Sigma}\bra{\udot-\:12\phi} +2\xi\Omega
+\delta_a\dotn^a +\udot^a\bra{\alpha_a-a_a}
+\bra{a^a-\udot^a}\bra{\Sigma_a-\lc_{ab}\Omega^b}
 -\zeta^{ab}\Sigma_{ab} +Q,\label{dotphinl}
\ee

$u^a\lc^{bc}R_{abc}$:
\ba
\dot\xi &=& \bra{\:12\Sigma-\:13\theta}\xi
+\bra{\udot-\:12\phi}\Omega
 +\:12\bra{\hatn^a+\udot^a}\bras{\Omega_a+ \lc_{ab}\bra{\dotn^b+\Sigma^b}}
+\:12\lc_{ab}\delta^a\dotn^b -\:12\lc_{ca}\zeta_{b}^{~c}\Sigma^{ab}
+\:12\H,\label{dotxinl}
\ea

$u^cR_{c\lb ab\rb}$:
\ba
\dot\zeta_{\lb ab\rb}&=&\bra{\:12\Sigma-\:13\theta}\zeta_{ab} +\Omega\lc_{c\lb
a}\zeta_{b\rb}^{~~c} +\bra{\udot-\:12\phi}\Sigma_{ab}-\xi\lc_{c\lb
a}\Sigma_{b\rb}^{~~c}- \zeta_{c\lb a}\Sigma_{b\rb}^{~~c} +\delta_{\lb
a}\alpha_{b\rb} \nonumber\\
&&+\bra{\udot_{\lb a}-a_{\lb a}}\alpha_{b\rb} -\bra{\udot_{\lb a}+a_{\lb
a}}\bra{\Sigma_{b\rb}-\lc_{b\rb d}\Omega^d} -\lc_{c\lb
a}\H_{b\rb}^{~~c},\label{dotzetanl}
\ea

Then a 1+1+2 decomposition of the standard 1+3 evolution equations gives us the remaining evolution equations, which can't be found from $R_{abc}$.

Vorticity evolution equation:
\ba
\dot\Omega&=&\:12\lc_{ab}\delta^a\udot^b+\udot\xi+\Omega\bra{\Sigma-\:23\theta}
    +\Omega_a\bra{\Sigma^a+\alpha^a}
\ea

Shear evolution:
\ba
\dot\Sigma_{\lb ab\rb}&=&\delta_{\lb a}\udot_{b\rb} +\udot_{\lb a}\udot_{b\rb}
    -\Sigma_{\lb a}\bras{\Sigma_{b\rb}+2\alpha_{b\rb}}-\Omega_{\lb a}\Omega_{b\rb}
    +\udot\zeta_{ab}-\bra{\:23\theta+\:12\Sigma}\Sigma_{ab}-\Sigma_{c\lb a}\Sigma_{b\rb}^{~~c}
    -\E_{ab}+\:12\Pi_{ab}
\ea

\subsection{Mixture of propagation and evolution:}

$u^a\n^bR_{ab\bar c}=\n^au^bR_{ab\bar c}$:
\ba
\hat\alpha_{\bar a}-\dot a_{\bar a}&=&
    -\bra{\:12\phi+\udot}\alpha_a-\xi\lc_{ab}\alpha^b
    +\bra{\:13\theta+\Sigma}\bra{\udot_a-a_a}
    +\bra{\:12\phi-\udot}\bra{\Sigma_a+\lc_{ab}\Omega^b}\nonumber\\
   && -\xi\bra{\lc_{ab}\Sigma^b-\Omega_a}
    +\zeta_{ab}\bra{-\alpha^b+\Sigma^b+\lc^{bc}\Omega_c}
    +\:12 Q_a-\lc_{ab}\H^b,\label{hatalphanl}
\ea

$u^a\n^bu^cR_{abc}=-\n^au^bu^cR_{abc}$:
\ba
\hat \udot-\:13\dot\theta-\dot\Sigma &=& -\udot^2+\bra{\:13\theta+\Sigma}^2
    -2\alpha_a\Sigma^a+\Sigma_a\Sigma^a-\Omega_a\Omega^a-a_a\udot^a+\lc_{ab}\alpha^a\Omega^b\nonumber\\&&
    +\:16\bra{\mu+3p-2\Lambda}+\E-\:12\Pi\label{redundant}
\ea

Raychaudhuri equation:
\ba
\hat\udot-\dot\theta&=&-\delta_a\udot^a-\bra{\udot+\phi}\udot+\bra{a_a-\udot_a}\udot^a
    +\:13\theta^2
    +\:32\Sigma^2-2\Omega^2+2\Sigma_a\Sigma^a\nonumber\\&&
    -2\Omega_a\Omega^a
    +\Sigma_{ab}\Sigma^{ab}
    +\:12\bra{\mu+3p}-\Lambda\label{ray}
\ea

Vorticity evolution:
\ba
\dot\Omega_{\bar
a}+\:12\lc_{ab}\hat\udot^b&=&-\bra{\:23\theta+\:12\Sigma}\Omega_a
    +\Omega\bra{\Sigma_a-\alpha_a}+\:12\xi\udot_a+\:12\lc_{ab}\bras{-\udot
    a^b+\delta^b\udot-\:12\phi\udot^b}\nonumber\\&&
    -\:12\lc_{ab}\zeta^{bc}\udot_c+\Sigma_{ab}\Omega^b
\ea

Shear evolution:
\ba
\dot\Sigma-\:23\hat\udot&=&\:13\bra{2\udot-\phi}\udot-\bra{\:23\theta+\:12\Sigma}\Sigma
    -\:23\Omega^2-\:13\delta_a\udot^a+\Sigma_a\bras{2\alpha^a-\:13\Sigma^a}\nonumber\\&&
    -\:13\udot_a\bras{2a^a-\udot^a}
    +\:13\Omega_a\Omega^a+\:13\Sigma_{ab}\Sigma^{ab}-\E+\:12\Pi\label{scalarshearev}
\ea
\ba
\dot\Sigma_{\bar a}-\:12\hat\udot_{\bar
a}&=&\:12\delta_a\udot+\bra{\udot-\:14\phi}\udot_a
    -\bra{\:23\theta+\:12\Sigma}\Sigma_a
    +\:12\udot a_a-\:32\Sigma\alpha_a-\Omega\Omega_a\nonumber\\&&
    -\:12\bra{\xi\lc_{ab}+\zeta_{ab}}\udot^b
    +\Sigma_{ab}\bra{\alpha^b-\Sigma^b}
    -\E_a+\:12\Pi_a
\ea

Energy conservation:
\ba
\dot\mu+\hat Q&=&-\delta_aQ^a-\theta\bra{\mu+p}-\bra{\phi+2\udot}Q -
    \:32\Sigma\Pi+\bra{a_a-2\udot_a}Q^a -2\Sigma_a\Pi^a-\Sigma_{ab}\Pi^{ab}
\ea

Momentum conservation:
\ba
\dot Q+\hat p+\hat\Pi&=&-\delta_a\Pi^a-\bra{\:32\phi+\udot}\Pi-\bra{\:43\theta+\Sigma} Q
    -\bra{\mu+p}\udot\nonumber\\&&
     +\bra{\alpha_a-\Sigma_a+\lc_{ab}\Omega^b}Q^a
    +\bra{2a_a-\udot_a}\Pi^a+\zeta_{ab}\Pi^{ab}
\ea

\ba
\dot Q_{\bar a}+\hat\Pi_{\bar a}&=&-\delta_ap+\:12\delta_a\Pi-\delta^b\Pi_{ab}
    -Q\bra{\alpha_a+\Sigma_a+\lc_{ab}\Omega^b}-\:32\Pi a_a
    -\bra{\:43\theta-\:12\Sigma}Q_a+\Omega\lc_{ab}Q^b\nonumber\\&&
    -\bra{\:32\phi+\udot}\Pi_a+\xi\lc_{ab}\Pi^b
    -\bra{\mu+p-\:12\Pi}\udot_a-\Sigma_{ab}Q^b-\zeta_{ab}\Pi^b
    +\Pi_{ab}\bra{a^b-\udot^b}
\ea

Electric Weyl evolution:

\ba
\dot\E+\:12\dot\Pi+\:13\hat Q&=&+\lc_{ab}\delta^a\H^c+\:16\delta_aQ^a
    +\bra{\:32\Sigma-\theta}\E
    -\:12\bra{\:13\theta+\:12\Sigma}\Pi +\:13\bra{\:12\phi-2\udot}Q
    +3\xi\H\nonumber\\&&
    -\:12\bra{\mu+p}\Sigma
    +\bra{2\alpha_a+\Sigma_a-\lc_{ab}\Omega^b}\E^a
    +\bra{\alpha_a-\:16\Sigma_a-\:12\lc_{ab}\Omega^b}\Pi^a\nonumber\\&&
    +\:13\bra{a_a+\udot_a}Q^a +2\lc_{ab}\udot^a\H^c
    -\Sigma_{ab}\bra{\E^{ab}+\:12\Pi^{ab}}
    +\lc_{ab}\H^{bc}\zeta^a_{~c}
\ea

\ba
\dot\E_{\bar a}+\:12\lc_{ab}\hat\H^b+\:12\dot\Pi_{\bar a}+\:14\hat Q_{\bar a} &=&
    \:34\lc_{ab}\delta^b\H+\:12\lc_{bc}\delta^b\H^c_{~a}-\:14\delta_a Q
    -\:12\bra{\mu+p-\:32\E+\:14\Pi}\Sigma_a\nonumber\\&&
    +\:34\bra{\E+\:12\Pi}\lc_{ab}\Omega^b
    -\:12Q\udot_a +\:32\H\lc_{ab}\udot^b
    -\:32\bra{\E+\:12\Pi}\alpha_a\nonumber\\&&
    -\:14Qa_a-\:34\H\lc_{ab}a^b
    +\bra{\:34\Sigma-\theta}\E_a -\:12\Omega\lc_{ab}\E^b
    +\:52\xi\H_a-\bra{\:14\phi+\udot}\lc_{ab}\H^b\nonumber\\&&
    +\:12\bra{\:14\phi-\udot}Q_a +\:14\xi\lc_{ab}Q^b
    -\:12\bra{\:13\theta+\:14\Sigma}\Pi_a-\:14\Omega\lc_{ab}\Pi^b\nonumber\\&&
    +\:12\Sigma_{ab}\bra{3\E^b-\:12\Pi^b}
    +\:12\bra{3\E_{ab}-\:12\Pi_{ab}}\Sigma^b
    -\bra{\E_{ab}+\:12\Pi_{ab}}\bra{\alpha^b+\:12\lc^{bc}\Omega_c}\nonumber\\&&
    +\:12\zeta_{ab}\bra{\lc^{bc}\H_c+Q^b}
    -\H_{ab}\lc^{bc}\udot_c
\ea

\ba
\dot\E_{\lb ab\rb}-\lc_{c\lb a}\hat\H_{b\rb}^{~~c}+\:12\dot\Pi_{\lb ab\rb}&=&
    -\lc_{c\lb a}\delta^c\H_{b\rb} -\:12\delta_{\lb a}Q_{b\rb}
    -\:12\bra{\mu+p+3\E-\:12\Pi}\Sigma_{ab}
    -\:12Q\zeta_{ab} -\:32\H\lc_{c\lb a}\zeta_{b\rb}^{~~c}\nonumber\\&&
    -\bra{\theta+\:32\Sigma}\E_{ab} +\Omega\lc_{c\lb a}\E_{b\rb}^{~~c}
    -\bra{\:16\theta-\:14\Sigma}\Pi_{ab}+\:12\Omega\lc_{c\lb a}\Pi_{b\rb}^{~~c}
    +\xi\H_{ab}\nonumber\\&&
    +\bra{\:12\phi+2\udot}\lc_{c\lb a}\H_{b\rb}^{~~c}
    -\udot_{\lb a}Q_{b\rb}
    -\bra{\alpha_{\lb a}+\:12\lc_{c\lb a}\Omega^c}\bra{2\E_{b\rb}+\Pi_{b\rb}}\nonumber\\&&
    +\Sigma_{\lb a}\bra{3\E_{b\rb}-\:12\Pi_{b\rb}}
    +2\lc_{c\lb a}\H_{b\rb}\bra{a^c-\udot^c}\nonumber\\&&
    +\Sigma_{c\lb a}\bra{3\E_{b\rb}^{~~c}-\:12\Pi_{b\rb}^{~~c}}
    +\lc_{c\lb a}\H_{b\rb d}\zeta^{cd}
\ea

Magnetic Weyl evolution:

\ba
\dot\H&=&-\lc_{ab}\delta^a\E^b+\:12\lc_{ab}\delta^a\Pi^b-3\xi\E
    +\bra{\theta+\:32\Sigma}\H+\Omega Q+\:32\xi\Pi
    -2\lc_{ab}\udot^a\E^b\nonumber\\&&
    +\bra{2\alpha_a+\Sigma_a-\lc_{ab}\Omega^b}\H^a
    -\:12\bra{\Omega_a+\lc_{ab}\Sigma^b}Q^a
    -\Sigma_{ab}\H^{ab}-\:12\lc_{ab}\E^{bc} \zeta^a_{~c}
\ea

\ba
\dot\H_{\bar a}-\:12\lc_{ab}\hat\E^b+\:14\lc_{ab}\hat\Pi^b&=&
    -\:34\lc_{ab}\delta^b\E+\:38\lc_{ab}\delta^b\Pi-\:12\lc_{bc}\delta^b\E^c_{~a}
    +\:14\lc_{bc}\delta^b\Pi^c_{~a}\nonumber\\&&
    +{\:34\H}\Sigma_a +{\:14Q}\lc_{ab}\Sigma^b
    +\:34Q\Omega_a+\:34\H\lc_{ab}\Omega^b
    -\:32\E\lc_{ab}\udot^b
    -\:32\H\alpha_a\nonumber\\&&
    +\:34\bra{\E-\:12\Pi}\lc_{ab}a^b
    -\:52\xi\E_a +\bra{\:14\phi+\udot}\lc_{ab}\E^b
    +\bra{\:34\Sigma-\theta}\H_a -\:12\Omega\lc_{ab}\H^b\nonumber\\&&
    +\:34\Omega Q_a-\:38\Sigma\lc_{ab}Q^b
    +\:54\xi\Pi_a-\:18\phi\lc_{ab}\Pi^b
    +\Sigma_{ab}\bra{\:32\H^b+\:14\lc^{bc}Q_c}\nonumber\\&&
    +\:32\lc_{ab}\zeta^{bc}\bra{\E_c-\:12\Pi_c+\:23\udot_c}
    +\H_{ab}\bra{\alpha^b+\:32\Sigma^b-\:12\lc^{bc}\Omega_c}
\ea

\ba
\dot\H_{\lb ab\rb}+\lc_{c\lb a}\hat\E_{b\rb}^{~~c}-\:12\lc_{c\lb a}\hat\Pi_{b\rb}^{~~c}&=&
    \lc_{c\lb a}\delta^c\E_{b\rb}-\;12\lc_{c\lb a}\delta^c\Pi_{b\rb}
    -\:32\H\Sigma_{ab}+\:12Q\lc_{c\lb a}\Sigma_{b\rb}^{~~c}\nonumber\\&&
    +\:32\bra{\E-\:12\Pi}\lc_{c\lb a}\zeta_{b\rb}^{~~c}
    -\xi\E_{ab}-\bra{\:12\phi+2\udot}\lc_{c\lb a}\E_{b\rb}^{~~c}\nonumber\\&&
    -\bra{\theta+\:32\Sigma}\H_{ab}-\Omega\lc_{c\lb a}\H_{b\rb}^{~~c}
    +\:12\xi\Pi_{ab}+\:14\phi\lc_{c\lb a}\Pi_{b\rb}^{~~c}\nonumber\\&&
    +\Sigma_{\lb a}\bra{3\H_{b\rb}-\lc_{b\rb c}Q^c}
    +\Omega_{\lb a}\bra{\:32Q_{b\rb}-\lc_{b\rb c}H^c}
    -2\alpha_{\lb a}\H_{b\rb}\nonumber\\&&
    +\E_{\lb a}2\lc_{b\rb c}\bra{a^c+\udot^c}
    -\Pi_{\lb a}\lc_{b\rb c}a^c
    +3\Sigma_{c\lb a}\H_{b\rb}^{~~c}
    -\lc_{c\lb a}\zeta^{cd}\bra{\E_{b\rb d}-\:12\Pi_{b\rb d}}
\ea

\subsection{Propagation equations}

Propagation and constraint equations are formed from either projecting $R_{abc}$ as indicated, or from projections of the 1+3 constraint equations, denoted $C_i$, as given in~\cite{HvE}.

$\n^a\N^{bc}R_{abc}$:
\ba
\hat\phi&=&-\:12\phi^2+2\xi^2+\bra{\:13\theta+\Sigma}\bra{\:23\theta-\Sigma}+\delta_aa^a-a_aa^a
    -\zeta_{ab}\zeta^{ab}+2\lc_{ab}\alpha^a\Omega^b-\Sigma_a\Sigma^a+\Omega_a\Omega^a\nonumber\\
    &&-\:23\bra{\mu+\Lambda}-\:12\Pi-\E,\label{hatphinl}
\ea

$\n^a\lc^{bc}R_{abc}$:
\ba
\hat\xi&=&-\phi\xi+\bra{\:13\theta+\Sigma}\Omega+\:12\lc_{ab}\delta^aa^b
    +\:12\lc_{ab}\Sigma^aa^b+\bra{\:12a_a+\alpha_a}\Omega^a\label{hatxinl}
\ea

$\n^aR_{a\lb bc\rb}$:
\ba
\hat\zeta_{\lb ab\rb}&=&-\phi\zeta_{ab}-\zeta^c_{~\lb a}\zeta_{b\rb c}+
\delta_{\lb a}\hatn_{b\rb }-\hatn_{\lb a}\hatn_{b\rb }+2\dotn_{\lb a}\lc_{b\rb c}\Omega^c
-\Omega_{\lb a}\Omega_{b\rb }-\Sigma_{\lb a}\Sigma_{b\rb }\nonumber\\
 &&+\bra{\:13\theta
+\Sigma}\Sigma_{ab}-\:12\Pi_{ab}-{\cal E}_{ab},\label{hatzetanl}
\ea

Shear divergence $(C_1)^a\n_a$:
\ba
\hat\Sigma-\:23\hat\theta&=&-\:32\phi\Sigma-2\xi\Omega-\delta_a\Sigma^a
    -\lc_{ab}\delta^a\Omega^b+2\Sigma_aa^a-2\lc_{ab}\udot^a\Omega^b
    +\Sigma_{ab}\zeta^{ab}-Q
\ea

and $(C_1)_{\bar a}$:
\ba
\hat\Sigma_{\bar a}-\lc_{ab}\hat\Omega^b&=&\:12\delta_a\Sigma
    +\:23\delta_a\theta - \lc_{ab}\delta^b\Omega-\:32\phi\Sigma_a
    +\xi\lc_{ab}\Sigma^b
    -\xi\Omega_a +\bra{\:12\phi+2\udot}\lc_{ab}\Omega^b\nonumber\\&&
    -\:32\Sigma a_a
    +\Omega\lc_{ab}\bras{a^b-2\udot^b}
    -\delta^b\Sigma_{ab}-\zeta_{ab}\Sigma^b+\Sigma_{ab}a^b+\lc_{ab}\zeta^{bc}\Omega_c
    -Q_a
\ea

Vorticity divergence equation ($C_2$):
\ba
\hat\Omega&=&-\delta_a\Omega^a+\bra{\udot-\phi}\Omega+\bra{a_a+\udot_a}\Omega^a\label{hatOmSnl}
\ea

$(C_3)_{\lb ab\rb}$:
\ba
\hat\Sigma_{\lb ab\rb}&=&\delta_{\lb a}\Sigma_{b\rb} -\lc_{c\lb a}\delta^c\Omega_{b\rb}
    -\:12\phi\Sigma_{ab}+\xi\lc_{c\lb a}\Sigma_{b\rb}^{~~c}
    +\:32\Sigma\zeta_{ab}-\Omega\lc_{c\lb a}\zeta_{b\rb}^{~~c}\nonumber\\&&
    -2\Sigma_{\lb a}a_{b\rb}-2\lc_{c\lb a}\udot^c\Omega_{b\rb}
    -\Sigma_{c\lb a}\zeta_{b\rb}^{~~c}-\lc_{c\lb a}\H_{b\rb}^{~~c}
\ea

Electric Weyl Divergence $(C_4)^a\n_a$:
\ba
\hat\E-\:13\hat\mu+\:12\hat\Pi&=&-\delta_a\E^a-\:12\delta_a\Pi^a
    -\:32\phi\bra{\E+\:12\Pi} +\bra{\:12\Sigma-\:13\theta} Q+3\Omega\H +\bra{2\E_a+\Pi_a}a^a\nonumber\\&&
    +\:12\Sigma_aQ^a+3\Omega_a\H^a-\:32\lc_{ab}\Omega^aQ^b+\lc_{ab}\Sigma^{ac}\H_c^{~b}
    +\bra{\E_{ab}+\:12\Pi_{ab}}\zeta^{ab}
\ea

$(C_4)_{\bar a}$:
\ba
\hat\E_{\bar a}+\:12\hat\Pi_{\bar a}&=& \:12\delta_a\E+\:13\delta_a\mu+\:14\delta_a\Pi
    -\delta^b\E_{ab}-\:12\delta^b\Pi_{ab}
    +\:12Q\Sigma_a+\H\lc_{ab}\Sigma^b
    -\:32\H\Omega_a-\:32Q\lc_{ab}\Omega^b\nonumber\\&&
    -\:32\bra{\E+\:12\Pi}a_a
    -\:32\phi\bra{\E_a+\:12\Pi_a}+\xi\lc_{ab}\bra{\E^b+\:12\Pi^b}
    +3\Omega\H_a-\Sigma\lc_{ab}\H^b\nonumber\\&&
    -\bra{\:13\theta+\:14\Sigma}Q_a +\:32\Omega\lc_{ab}Q^b
    +\:12\Sigma_{ab}Q^b
    -\zeta_{ab}\bra{\E^b+\:12\Pi^b}
    +\bra{\E_{ab}+\:12\Pi_{ab}}a^b
    +3\H_{ab}\Omega^b
\ea

Magnetic Weyl divergence $(C_5)^a\n_a$:
\ba
\hat\H&=& -\delta_a\H^a-\:12\lc_{ab}\delta^aQ^b
    -\:32\phi\H-\bra{3\E+\mu+p-\:12\Pi}\Omega-Q\xi\nonumber\\&&
    +2\H_aa^a-3\Omega_a\bra{\E^a-\:16\Pi^a}
    +\zeta_{ab}\H^{ab}-
    \lc_{ab}\Sigma^a_{~c}\bra{\E^{bc}+\:12\Pi^{bc}}
\ea

$(C_5)_{\bar a}$:
\ba
\hat\H_{\bar a}-\:12\lc_{ab}\hat Q^b &=& \:12\delta_a\H-\delta^b\H_{ab}
    -\:12\lc_{ab}\delta^b Q
    -\:32\bra{\E+\:12\Pi}\lc_{ab}\Sigma^b
    -\bra{-\:32\E+\mu+p+\:14\Pi}\Omega_a\nonumber\\&&
    -\:32\H a_a+\:12Q\lc_{ab}a^b
    -3\Omega\E_a+\:32\Sigma\lc_{ab}\E^b
    -\:32\phi\H_a+\xi\lc_{ab}\H^b\nonumber\\&&
    -\:12\xi Q_a+\:14\phi\lc_{ab}Q^b
    +\:12\Omega\Pi_a+\:34\Sigma\lc_{ab}\Pi^b\nonumber\\&&
    +\H_{ab}a^b-\zeta_{ab}\H^b-3\bra{\E_{ab}-\:16\Pi_{ab}}\Omega^b
    +\:12\lc_{ab}\zeta^{bc}Q_c
\ea

\subsection{Constraint}

$\lc^{ab}u^cR_{abc}$:
\ba
\delta_a\Omega^a+\lc_{ab}\delta^a\Sigma^b&=&\bra{2\udot-\phi}\Omega
    -3\xi\Sigma
    +\lc_{ab}\zeta^{ac}\Sigma^b_{~c} +\H\label{divOmeganl}
\ea

$\N^{bc}R_{\bar abc}$:
\ba
\:12\delta_a\phi-\lc_{ab}\delta^b\xi-\delta^b\zeta_{ab}&=&-2\xi\lc_{ab}a^b
    -\Omega\bra{\Omega_a+\lc_{ab}\Sigma^b-2\lc_{ab}\alpha^b}
    -\bra{\:13\theta-\:12\Sigma}\bra{\Sigma_a-\lc_{ab}\Omega^b}\nonumber\\&&
    -\bra{\Sigma^b-\lc^{bc}\Omega_c}\Sigma_{ab}
    -\:12\Pi_a-\E_a\label{divzetanl}
\ea

From $(C_3)_{ab}\n^b$ and $(C_1)_{\bar a}$, or $\n^a u^cR_{a\bar bc}$
\ba
\delta_a\Sigma-\:23\delta_a\theta+2\lc_{ab}\delta^b\Omega +2\delta^b\Sigma_{ab}&=&
    -\phi\bra{\Sigma_a-\lc_{ab}\Omega^b}-2\xi\bra{\Omega_a-3\lc_{ab}\Sigma^b}
    -4\Omega\lc_{ab}\udot^b\nonumber\\&&
    +2\zeta_{ab}\Sigma^b+2\lc_{ab}\zeta^{bc}\Omega_c
    +\Sigma_{ab}a^b-2\lc_{ab}\H^b-Q_a
\ea

Finally, we note that the equation formed from
$(C_3)_{ab}\n^a\n^b$ is equivalent to Eqs~(\ref{hatOmSnl})
and~(\ref{divOmeganl}).

It is worth noting that one of Eqs.~(\ref{redundant}),~(\ref{ray}),~(\ref{scalarshearev}) is redundant since (\ref{redundant})$=\:13$(\ref{ray})$-$(\ref{scalarshearev}). Also, note that there are no evolution equations for $\udot, \udot_a, \alpha_a$, and there is no propagation equation for $a_a$; these all must be determined by specifying a choice of frame.

\subsection{Maxwell's Equations}

For completeness we also give the decomposition of Maxwell's
equations, previously given in~\cite{CMBD}.
We decompose the electric and magnetic field vectors as
\ba
E^a&=&\El\n^a+\El^a,\\
B^a&=&\B \n^a+\B^a,
\ea
while the 3-current may be written as
\be
j^a={\cal J}\n^a+{\cal J}^a.
\ee
Maxwell's equations then become:
\ba
 \hat \El +\delta_a \El ^a &=&   -\phi \El  + \El _a\hatn^a +2\Omega \B
+2\Omega^a \B _a + \mu_0 \rho_{\mathrm{e}},\\
 \hat \B +\delta_a \B ^a &=&   -\phi \B  + \B _a\hatn^a -2\Omega \El
-2\Omega^a \El _a,\label{MEgen1}\\
 \dot \El -\lc_{ab}\delta^a \B ^b &=&   2\xi
\B +\El ^a\dotn_a-\bra{\tfrac{2}{3}\theta-\Sigma} \El
+\Sigma^a \El _a +\lc_{ab}\bra{\udot^a \B ^b +\Omega^a \El ^b}-
\mu_0{\cal J},\\
 \dot \B +\lc_{ab}\delta^a \El ^b &=&-2\xi
\El +\B ^a\dotn_a-\bra{\tfrac{2}{3}\theta-\Sigma} \B
+\Sigma^a \B _a -\lc_{ab}\bra{\udot^a \El ^b -\Omega^a \B ^b},\\
 \dot \El _{\bar a}+\lc_{ab}\bra{\hat \B ^b-\delta^b \B } &=&    \xi \B _a
-\bra{\tfrac{1}{2}\phi+\udot}\lc_{ab}\B ^b
-\bra{\tfrac{2}{3}\theta+\tfrac{1}{2}\Sigma}\El _a-\Omega\lc_{ab}\El
^b\nonumber\\&&
 + \El \bra{-\dotn_{a}+\Sigma_a+\lc_{ab}\Omega^b} +\B
\lc_{ab}\bra{\udot^b-\hatn^b}
+\Sigma_{ab}\El ^b-\lc_{ab}\zeta^{bc}\B _c-\mu_0{\cal J}_a,\label{MEgen5}\\
 \dot \B _{\bar a}-\lc_{ab}\bra{\hat \El ^b-\delta^b \El } &=& -\xi \El _a
+\bra{\tfrac{1}{2}\phi+\udot}\lc_{ab}\El ^b
-\bra{\tfrac{2}{3}\theta+\tfrac{1}{2}\Sigma}\B _a-\Omega\lc_{ab}\B
^b\nonumber\\&&
 + \B \bra{-\dotn_{a}+\Sigma_a+\lc_{ab}\Omega^b} -\El
\lc_{ab}\bra{\udot^b-\hatn^b}
+\Sigma_{ab}\B ^b+\lc_{ab}\zeta^{bc}\El _c.\label{MEgen}
\ea
Here, MKS units are used ($\mu_0$), and $\rho_e$ is the charge density. The first two equations arise from the constraint ME,
while the rest
are the evolution ME. In flat space in the absence of currents and
charges the rhs of these equations vanish (for a static `natural' choice of
frame). Thus, gravity modifies ME in the form of generalised currents. Note
how the rotation terms $\xi,~\Omega$ and $\Omega^a$ flip the parities of the
EM fields.

\section{Perturbations of spherically symmetric and LRS spacetimes}

The utility of the approach presented here is that for LRS spactimes, for which all quantities are rotationally symmetric about a preferred spatial direction (i.e., they admit a one-dimensional isotropy group), all the non-zero 1+1+2 variables are scalars. This direction may be specified, for example, by a non-degenerate eigenvector of the electric Weyl tensor, or by the vorticity vector. A full discussion of LRS spacetimes in the covariant approach is given in~\cite{vEE}; see their Table~1 for a summary of the different cases which can occur, in a notation similar to that presented here. 

The fact that background quantities are scalars in LRS spacetimes means that under linear perturbations, all vector and tensor quantities are automatically gauge invariant, by the Stewart-Walker Lemma~\cite{SW}. We shall now give an overview of how to set up the perturbation equations.

In the background, which we shall take as a general LRS spacetime, all vector and tensor equations are automatically zero, resulting in the set
\ba
\dot\phi &=&
\bra{\:23\theta-\Sigma}\bra{\udot-\:12\phi} +2\xi\Omega+Q,
\\
\dot\xi &=& \bra{\:12\Sigma-\:13\theta}\xi
+\bra{\udot-\:12\phi}\Omega+\:12\H,
\\
\dot\Omega&=&+\udot\xi+\Omega\bra{\Sigma-\:23\theta},
\\
\hat\udot-\dot\theta&=&-\bra{\udot+\phi}\udot+\:13\theta^2
    +\:32\Sigma^2-2\Omega^2
    +\:12\bra{\mu+3p}-\Lambda,
\\
\dot\Sigma-\:23\hat\udot&=&\:13\bra{2\udot-\phi}\udot-\bra{\:23\theta+\:12\Sigma}\Sigma
    -\:23\Omega^2-\E+\:12\Pi,
\\
\dot\mu+\hat Q&=&-\theta\bra{\mu+p}-\bra{\phi+2\udot}Q -
    \:32\Sigma\Pi,
\\
\dot Q+\hat p+\hat\Pi&=&-\bra{\:32\phi+\udot}\Pi-\bra{\:43\theta+\Sigma} Q
    -\bra{\mu+p}\udot,
\\
\dot\E+\:12\dot\Pi+\:13\hat Q&=&
    +\bra{\:32\Sigma-\theta}\E
    -\:12\bra{\:13\theta+\:12\Sigma}\Pi\nonumber\\&&
    +\:13\bra{\:12\phi-2\udot}Q
    +3\xi\H
    -\:12\bra{\mu+p}\Sigma,
\\
\dot\H&=&-3\xi\E
    +\bra{\theta+\:32\Sigma}\H+\Omega Q+\:32\xi\Pi,
\\
\hat\phi&=&-\:12\phi^2+2\xi^2+\bra{\:13\theta+\Sigma}\bra{\:23\theta-\Sigma}
    -\:23\bra{\mu+\Lambda}-\:12\Pi-\E,\label{hatphinl}
\\
\hat\xi&=&-\phi\xi+\bra{\:13\theta+\Sigma}\Omega,
\\
\hat\Sigma-\:23\hat\theta&=&-\:32\phi\Sigma-2\xi\Omega-Q,
\\
\hat\Omega&=&+\bra{\udot-\phi}\Omega\label{hatOmSnl},
\\
\hat\E-\:13\hat\mu+\:12\hat\Pi&=&
    -\:32\phi\bra{\E+\:12\Pi} +\bra{\:12\Sigma-\:13\theta}Q+3\Omega\H,
\\
\hat\H&=&
    -\:32\phi\H-\bra{3\E+\mu+p-\:12\Pi}\Omega-Q\xi,
\\
0&=&\bra{2\udot-\phi}\Omega
    -3\xi\Sigma+\H.
\ea
These equations were first presented in this form in~\cite{BC} for LRS Class II models (which satisfy $\xi=\Omega=0\Rightarrow\H=0$, and were shown to be consistent with the commutation relation~(\ref{comm-un}). 

It is perhaps easier to think of these in matrix form. Let
\be
{\bm X}=\left(
\begin{array}{c}
  \phi \\
  \theta \\
  \Sigma \\
  {\cal A} \\
  \Omega \\
  \xi \\
  \mathcal{E} \\
  \mathcal{H} \\
  \mu \\
  p \\
  Q \\
  \Pi
\end{array}
\right)
\ee
be the column matrix of all non-zero scalar quantities. Depending on the LRS model in question $\bm X$ will not be this big. For example, for the Schwarzschild solution we have just $\bm X=(\phi,\mathcal{A,E})^T$.  
Then, in general, this system of equations may be cast in the form
\be
\bm\alpha\dot{\bm X}+\bm\beta\hat{\bm X}=\bm\Gamma\bm X+\bm X^T\bm\Delta\bm X\label{matrix}
\ee
where the constant matrices $\bm\alpha, \bm\beta, \bm\Gamma, \bm\Delta$ may be read off from the above equations. 

We can now set up the perturbative proceedure schematically as follows:
\begin{enumerate}
\item \emph{Find a complete set of gauge-invariant perturbation variables.} This may be achieved by defining
\be
\bm \Psi_a=\delta_a\bm X\ ;
\ee 
i.e., by taking angular derivatives of the background variables we find a new set of gauge-invariant variables. The remaining GI variables are all the 1+1+2 vectors and tensors: $\bm\chi_a=(\mathcal{E}_a,a_a,\ldots),\bm\chi_{ab}=(\zeta_{ab},\mathcal{E}_{ab},\mathcal{H}_{ab},\ldots)$, which obey linearised versions of the above 1+1+2 equations.   Under perturbations Eq.~(\ref{matrix}) becomes
\be
\bm\alpha\dot{\bm X}+\bm\beta\hat{\bm X}=\bm\Gamma\bm X+\bm X^T\bm\Delta\bm X+\bm A\delta^a\bm\chi_a+\bm B\varepsilon_{ab}\delta^a\bm\chi^b\label{matrix2},
\ee
where the matrices $\bm A, \bm B$ have constant coefficients.  
 Evolution and propagation equations for the new GI variables $\bm\Psi_a$ may be found by taking the angular derivative of  Eq.~(\ref{matrix2}), and using the commutation relations~(\ref{commdd}) and~(\ref{commdh}),
 giving: 
\ba
\bm\alpha\dot{\bm \Psi}_a+\bm\beta\hat{\bm \Psi}_a&=&\left[\bm\Gamma+\left(\:12\Sigma-\:13\theta-\:12\phi\right)\bm\alpha\right]\bm\Psi_a
-\left(\Omega\bm\alpha+\xi\bm\beta\right)\varepsilon_{ab}\bm\Psi^b\nonumber\\&&
+\bm X^T\bm\Delta\bm \Psi_a++\bm \Psi_a^T\bm\Delta\bm X +\bm A\delta_a\delta^b\bm\chi_b+\bm B\varepsilon_{bc}\delta_a\delta^b\bm\chi^c.
\ea
These equations replace the corresponding system~(\ref{matrix2}) in the 1+1+2 equations.

\item \emph{Harmonic analysis} Two parities of harmonics may be introduced, generalising the axial and polar modes for spherical symmetry. These were first defined in~\cite{CB,BC}, and are discussed in Appendix~\ref{harmonics}. These are analogous to the scalar-vector-tensor decomposition in FLRW models. After this, all variables become scalars, which are functions of two affine parameters associated with $u^a$ and $n^a$. 

\item \emph{Master Variables} At this stage the governing system of equations is linear in the perturbation variables $\bm\Phi$ and $\bar{\bm\Phi}$, which are the column vectors containing all the even and odd harmonically decomposed variables, and splits into two parities. We then have two linear systems of equations looking like
\be
\bm\gamma \dot{\bm\Phi}+\bm\lambda\hat{\bm\Phi}=\bm \Xi\bm\Phi 
\ee
where $\bm\Xi$ is a matrix with coefficients depending only on the background parameters (as well as the harmonic index $k$), and $\bm\gamma,\bm\lambda$ are constant matrices. The true degrees of freedom of this system will be governed by a reduced set of frame independent master variables, which will obey a closed set of wave equations. Finding these can be tricky. All other variables are related to the master variables by quadrature, plus frame degrees of freedom. See~\cite{CB} for the full details in the Schwarzschild case.  

\end{enumerate}

These are the key steps required given a particular LRS model is chosen. Steps 1 and 2 are algorithmic; step 3 can be very difficult.

\section{summary}

We have presented a new semi-tetrad approach to analysing Einstein's field equations. By introducing a single space-like vector into the 1+3 approach we decomposed the 1+3 equations into a system of evolution, propagation and constraint equations.  These were supplemented by a 1+1+2 decomposition of the Ricci equations for the spatial vector. Although presented in restricted form elsewhere, the full system was presented here for the first time.  

A key feature of the approach is that under a complete decomposition all objects are covariantly defined scalars, 2-vectors in the sheet and transverse-traceless 2-tensors, also in the sheet.  In an LRS spacetime, provided the spatial vector is chosen appropriately, all the vectors and tensors vanish, leaving just scalars. Under perturbations all indexed objects are first-order ensuring that there are no tensorial products; this ensures that we can introduce natural harmonic functions on the background which remove all tensorial properties of the equations. Finally, we are left with a system of gauge-invariant and covariant first-order PDEs to manipulate. The solution of this system provides the solution of the perturbation problem.

\acknowledgements The author wishes to thank Peter Dunsby, George Ellis, Roy Maartens, and Mattias Marklund, for many useful discussions (though some time ago now...), and in particular wishes to thank Richard Barrett and Gerold Betschart for the same, as well as checking many of the equations, and GB for supplying Eqs.~(\ref{commv-un})~-- (\ref{commv1}). CC is funded by the NRF (South Africa).

\appendix

\section{Useful relations for decomposing equations}\label{split}

Given any 1+3 vectors and tensors, we may decompose them as
\ba
x^a&=&X\n^a+X^a,\\
y^a&=&Y\n^a+Y^a,\\
\psi_{ab}&=&\psi_{\<ab\>}=\Psi\bra{\n_a\n_b-\:12\N_{ab}}+2\Psi_{(a}\n_{b)}+\Psi_{{ab}},\\
\phi_{ab}&=&\phi_{\<ab\>}=\Phi\bra{\n_a\n_b-\:12\N_{ab}}+2\Phi_{(a}\n_{b)}+\Phi_{{ab}}.
\ea
Then we have the following expansions from 1+3 quantities $\longrightarrow$ 1+1+2 variables:
\ba
x_ax^a&=&X^2+X_aX^a,\\
\eta_{abc}x^by^c&=& \bra{\lc_{bc}X^bY^c}\n_a+\lc_{ab}\bra{YX^b-XY^b},\\
x_{\<a}y_{b\>}&=&\:13\bra{2XY-X_cY^c}\bra{\n_a\n_b-\:12\N_{ab}}
    +\bras{XY_{(a}+YX_{(a}}\n_{b)}
    +X_{\lb a}Y_{b\rb}, \\
\psi_{ab}x^b&=&\bra{X\Psi+X_b\Psi^b}\n_a-\:12\Psi X_a+X\Psi_a+\Psi_{ab}X^b,\\
\eta_{cd\<a}x^c\psi_{b\>}^{~~d}&=&\lc_{cd}X^c\Psi^d\bra{\n_a\n_b-\:12\N_{ab}}+
    \bras{\bra{X\Psi^c-\:32\Psi X^c}\lc_{c(a} +\lc_{cd}X^c\Psi^d_{~(a}}\n_{b)}
    +X\lc_{c\lb a}\Psi_{b\rb}^{~~c}-X^c\lc_{c\lb a}\Psi_{b\rb},\\
\psi_{ab}\psi^{ab}&=&\:32\Psi^2+2\Psi_a\Psi^a+\Psi_{ab}\Psi^{ab},\\
\psi_{c\<a}\phi_{b\>}^{~~c}&=& \bra{\:12\Psi\Phi+\:13\Psi_c\Phi^c-\:13\Psi_{cd}\Phi^{cd}}
    \bra{\n_a\n_b-\:12\N_{ab}}
    +\bras{\:12\Psi\Phi_{(a}+\:12\Phi\Psi_{(a}+\Psi^c\Phi_{c(a}+\Phi^c\Psi_{c(a}}\n_{b)}\nonumber\\&&
    -\:12\Psi\Phi_{ab}-\:12\Phi\Psi_{ab}+\Psi_{\lb a}\Phi_{b\rb}+\Psi_{c\lb
    a}\Phi_{b\rb}^{~~c},\\
\eta_{abc}\psi^b_{~d}\phi^{dc}&=&\n_a\lc_{bc}\Psi^b_{~d} \Phi^{dc}
    +\:32\lc_{ab}\bra{\Phi\Psi^b-\Psi\Phi^b}.
\ea
For 1+3 derivatives we find:
\ba
\dot x_{\< a\>}&=& \bra{\dot X-X_b\alpha^b}\n_a+X\alpha_a+\dot X_{\bar a},\\
\dot\psi_{\<ab\>}&=&
\bra{\dot\Psi-2\Psi_c\alpha^c}\n_a\n_b
    -\:12\dot\Psi\N_{ab}
    +\bras{3\Psi\alpha_{(a}+2\dot\Psi_{(\bar a}-2\alpha^c\Psi_{c(a}}\n_{b)}
    +2\Psi_{(a}\alpha_{b)}+\dot\Psi_{\lb ab\rb},\\
\sdel_ax^a&=&\hat X+X\phi-X_aa^a+\delta_aX^a,\\
\eta_{abc}\sdel^bx^c&=& \bra{2X\xi+\lc_{bc}\delta^bX^c}\n_a+\xi X_a
    +\lc_{ab}\bras{-Xa^b+\delta^b X -\hat X^b-\:12\phi X^b-\zeta^{bc}X_c},\\
\sdel_{\<a}x_{b\>}&=&\:13\bras{2\hat X-\phi
    X-2X_ca^c-\delta_cX^c}\bra{\n_a\n_b-\:12\N_{ab}}\nonumber\\&&+
    \bras{Xa_{(a}+\delta_{(a}X+\hat X_{(\bar a}-\:12\phi
    X_{(a}+X^c\bra{\xi\lc_{c(a}-\zeta_{c(a}}}\n_{b)}
    +X\zeta_{ab}+\delta_{\lb a}X_{b\rb},\\
\sdel^b\psi_{ab}&=&
\bra{\hat\Psi+\:32\phi\Psi-2\Psi_ba^b+\delta_b\Psi^b-\Psi_{bc}\zeta^{bc}}\n_a
    +\hat\Psi_{\bar a}+\:32\phi\Psi_a+\:32\Psi
    a_a-\:12\delta_a\Psi\nonumber\\&&
    -\Psi_{ab}a^b+\bras{-\xi\lc_{ab}+\zeta_{ab}}\Psi^b+\delta^b\Psi_{ab},\\
\eta_{cd\<a}\sdel^c\psi_{b\>}^{~~d}&=& \bra{3\xi\Psi+\lc_{cd}\delta^c\Psi^d-\lc_{cd}\Psi^{de}\zeta^c_{~e}}
    \bra{\n_a\n_b-\:12\N_{ab}}\nonumber\\&&
    +\brac{ \bras{-\:32\delta^c\Psi+\:32\Psi a^c+\hat\Psi^c+\:12\phi\Psi^c
    +2\Psi_d\zeta^{cd}}\lc_{c(a}
    +5\xi\Psi_{(a} +\lc^{cd}\bras{\Psi_d\zeta_{c(a}+\delta_c\Psi_{d(a}
    }}\n_{b)}\nonumber\\&&
    -\lc_{c\lb a}\delta^c\Psi_{a\rb} +2\lc_{c\lb a}{a^c\Psi_{b\rb}}
    +\lc_{c\lb a}\hat\Psi^c_{~b\rb}
    +\:12\phi\lc_{c\lb a}\Psi^c_{~b\rb}-\:32\Psi\lc_{c\lb a}\zeta^c_{~b\rb}
    +\xi\Psi_{ab}+\lc_{c\lb a}\Psi_{b\rb d}\zeta^{cd}.
\ea
Given any relation in 1+3 notation, these relations may be substituted directly to aid decomposition.

\section{Harmonic Functions}\label{harmonics}

We introduce dimensionless harmonic functions $Q$, defined on any LRS background, as
eigenfunctions of the 2-dimensional Laplace-Beltrami operator:
\be
\delta^2 Q = -\frac{k^2}{r^{2}} Q, \qquad\hat Q=0=\dot Q \qquad (0
\leq k^2).\label{SH}
\ee
The function $r$ is, up to an irrelevant constant, covariantly
defined by
\be
\frac{\hat r}{r}\, \equiv \frac12\phi, \qquad  \frac{\dot r}{r} \equiv
\frac13\theta-\frac12\Sigma, \qquad \delta_a r \equiv 0. \label{rdef}
\ee
While we haven't chosen a specific basis for $Q$, we can now expand any first
order scalar ${\psi}$ in terms of these functions schematically as
\be
{\psi}=\sum_{k}{\psi}_{\Si}^{(k)} Q^{(k)} = {\psi}_{\Si} Q,
\ee
where the sum (or integral) over $k$  is implicit in the last equality. The
$\Si$ subscript reminds us that ${\psi}$ is a scalar, and that a
harmonic expansion has been made.

We also need to expand vectors in  harmonics. We therefore define
the \emph{even} (electric) parity vector  harmonics as
\be
Q_a^{(k)}=r\delta_a Q^{(k)} ~~~\Rightarrow ~~~\hat Q_{\bar
a}=0=\dot Q_{\bar a},~~~\delta^2Q_a=\bra{1-k^2}r^{-2}Q_a;
\ee
where the $(k)$ superscript is implicit, and we define \emph{odd}
(magnetic) parity vector harmonics as
\be
\bar Q_a^{(k)}=r\lc_{ab}\delta^b Q^{(k)}~~~\Rightarrow
~~~\hat{\bar{Q}}_{\bar a}=0=\dot{\bar{Q}}_{\bar a},~~~\delta^2\bar
Q_a=\bra{1-k^2}r^{-2}\bar Q_a.
\ee
Note that $\bar Q_a=\lc_{ab}Q^b\Leftrightarrow Q_a=-\lc_{ab}\bar
Q^b$, so that $\lc_{ab}$ is a parity operator. The crucial
difference between these two types of vector harmonics is that
$\bar Q_a$ is solenoidal, so
\be
\delta^a\bar Q_a=0,
\ee
while
\be
\delta^aQ_a=-k^2r^{-1} Q.
\ee
Note also that
\be
\lc_{ab}\delta^a  Q^b=0,~~~\mbox{and}~~~\lc_{ab}\delta^a\bar
Q^b=+k^2r^{-1} Q.
\ee
The harmonics are orthogonal: $Q^a\bar Q_a=0$ (for each $k$),
which implies that any first-order vector ${\psi}_a$ can now be
written
\be
{\psi}_a=\sum_{k} {\psi}^{(k)}_{\Vi} Q_a^{(k)}+\bar
{\psi}^{(k)}_{\Vi}\bar Q_a^{(k)}={\psi}_{\Vi} Q_a+\bar
{\psi}_{\Vi}\bar Q_a.
\ee
Again, we implicitly assume a sum over $k$ in the last equality,
and the $\Vi$ subscript reminds us that ${\psi}_a$ is a vector
expanded in harmonics.

Similarly we define even and odd tensor spherical harmonics as
\ba
Q_{ab}&=& r^2\delta_{\lb a}\delta_{b\rb}Q,~~~\Rightarrow ~~~\hat Q_{ab}=0=\dot
Q_{ab},
\\
\bar Q_{ab}&=& r^2\lc_{c\lb a}\delta^c\delta_{b\rb}Q,~~~\Rightarrow ~~~\hat{\bar
Q}_{ab}=0=\dot{\bar Q}_{ab},
\ea
which are orthogonal: $Q_{ab}\bar Q^{ab}=0$, and are parity inversions of one
another: $Q_{ab}=-\lc_{c\lb a}\bar Q_{b\rb}^{~~c}\Leftrightarrow
\bar Q_{ab}=\lc_{c\lb a} Q_{b\rb}^{~~c}$. Any first-order tensor may be expanded
\be
{\Psi}_{ab}=\sum_{k}^{} {\Psi}\T^{(k)}Q_{ab}^{(k)}+\bar
{\Psi}\T^{(k)}\bar Q_{ab}^{(k)}={\Psi}\T Q_{ab}+\bar {\Psi}\T\bar Q_{ab}.
\ee

\end{document}